# Metastable Staus: Reconstructing Non-Prompt Tracks at the ILC with the SiD Detector

Christopher Betancourt, F. Alexander Bogert and Bruce A. Schumm

Santa Cruz Institute for Particle Physics and the University of California at Santa Cruz – Department of Physics
1156 High Street, Santa Cruz, CA 95062 – USA

We have explored the reconstruction of kinks from the decay of metastable staus into a gravitino and a stau lepton in the radial region between the outer layer of the vertex detector and the second central tracking layer. After applying a cut on track multiplicity, we find a high efficiency for such events, with minimal background from Standard Model processes.

With its small number of precise tracking layers, the question of reconstruction non-prompt tracks has been an ongoing concern for the SiD Detector. In this paper, we report on the status of a study of the reconstruction of the decay $\tilde{\tau} \to \tau \, \tilde{g}$, for which the stau is metastable, decaying in the radial region that lies between the outer layer of the vertex detector and the second central tracking region. Here, $\tilde{\tau}$ represents the Supersymmetric partner of the tau lepton (stau), and $\tilde{g}$ the Supersymmetric partner of the graviton (gravitino). Decays of this sort arise naturally within the Gauge Mediated model of Supsersymetry breaking (GMSB) [1]; the relatively weak coupling between the SU(3)xSU(2)xU(1) and gravity sectors that produces the macroscopic decay lengths of interest lies within the cosmologically-allowed region.

Using the GEANT-based [2] SiD detector simulation, we have generated a sample of 1000 $e^+ e^- \to \tilde{\tau}^+ \tilde{\tau}^-$ events, with the stau decaying to tau-gravitino with a branching fraction of 100%. The stau-production events have been generated at a center-of-mass energy of 500 GeV, producing staus of mass 75 GeV/$c^2$. The cross section for this process is 90 fb, leading to an integrated signal Monte Carlo luminosity of 11.1 fb$^{-1}$. The coupling to the gravitational sector has been chosen so that the mean decay length of the staus is $\beta\gamma c\tau$ = 23 cm, approximately the radius of the first central-tracking layer. Backgrounds are estimated with a sample of 5.3 fb$^{-1}$ of combined Standard-Model processes.

Standard SiD track-finding algorithms are used to reconstruct charged-particle trajectories in the signal and background events. Tracks originating within the beampipe, and traveling at least through the outer layer of the vertex detector ("inner" tracks) are reconstructed with the SiD SeedTracker algorithm, while tracks originating between the outer layer of the vertex detector and the second tracking layer, and which subsequently travel into the calorimeter ("outer" tracks), are reconstructed with both the SeedTracker and Garfield algorithms. Inner and outer tracks are combined into "kinked tracks" according to the following criteria:

- The outer hit of the inner track must be on the outer layer of the vertex detector or the inner layer of the central tracker;

- The inner layer of the outer track must be outside the outer layer of the inner track, with no more than one missing layer;
- The inner and outer tracks must be on the same side of the barrel in z;
- The inner and outer tracks must have an intersection point in the x-y plane.

In addition, when the inner track has at least one central-tracker hit:

- The inner and outer curvature signs must match;
- The inner curvature must be less than the outer curvature.

The use of this additional information allows for somewhat better performance of the kink-finding algorithm for staus that decay between the first and second central-tracking layers.

Fig. 1 shows the reconstruction efficiency for stau-decay kinks as a function of the radius of the stau decay point. Stau decays were only considered if the stau track lay within $|\cos\theta| < 0.5$, and if the subsequent tau decay was a single-prong decay. Stau-decay kinks were only considered to be successfully reconstructed if they satisfied strict truth-matching criteria: each hit on the inside (outside) track was required to originate from the stau (charged tau decay) particle, while the reconstructed kink radius was required to be within 1 mm of the true stau decay point.

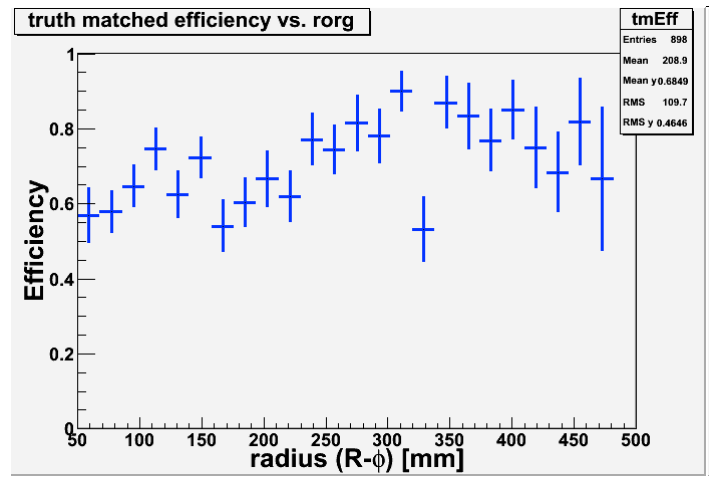

Figure 1: Stau kink reconstruction efficiency as a function of stau decay radius. Only one-prong tau decays are considered. A strict truth-matching requirement is placed on the reconstructed kinks (see text).

Fig. 2 shows the relative population of signal to background after kink reconstruction, again as a function of kink radius, scaled to an integrated luminosity of 10 fb$^{-1}$ at 500 GeV. The rate of kink reconstruction from Standard Model processes dominates that expected from this new physics scenario, particularly at low radius. However, there are a number of other characteristics that can be used to distinguish Standard Model events from those of greater interest. In particular, Fig. 3 shows the resulting distribution after a requirement that there be no greater than two reconstructed inside tracks.

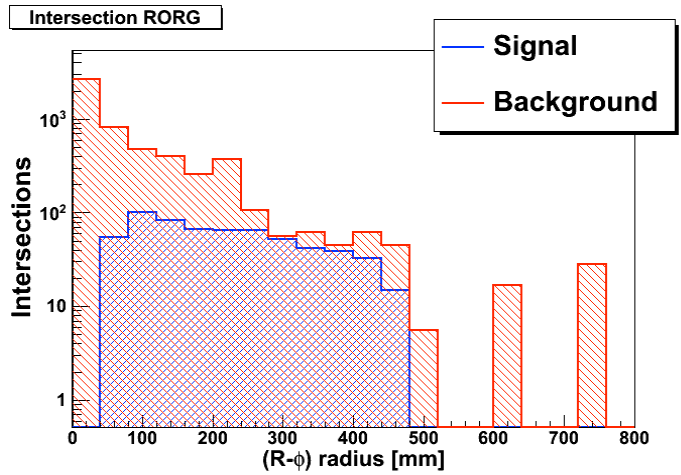

Figure 2: Distribution of reconstructed kinks, as a function of kink radius, for truth-matched signal and Standard-Model background, per 10 fb$^{-1}$ integrated luminosity.

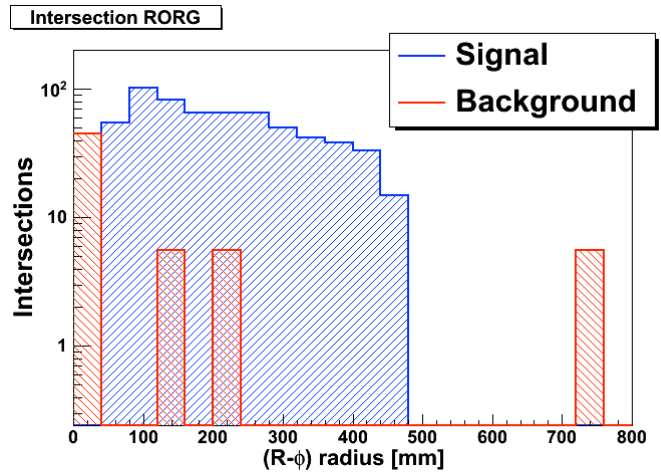

Figure 3: Distribution of reconstructed kinks, as a function of kink radius, for truth-matched signal and Standard-Model background, after a requirement that there be no greater than two reconstructed inside tracks.

After the limitation on the number of tracks is applied, the background is essentially eliminated. Thus, it seems relatively straightforward to reconstruct the metastable stau signature with the SiD detector for decay lengths between 6 cm (the outer radius of the vertex detector) and 48 cm (the radius of the second layer of the central tracker). It should be pointed out that, having the same proposed vertex detector, the ILD detector is unlikely to be able to reconstruct kinks at smaller radius than this, but it may be possible for the ILD to reconstruct higher-radius kinks. To this end, we are working on the development of a tracking algorithm that can reconstruct kinks using just three central tracking hits (plus calorimeter information). If successful, this would allow kinks to be reconstructed to a radius of 72cm with the SiD detector.

## Acknowledgments

The authors acknowledge and are grateful for the support provided by the United States Department of Energy, under contracts DE-SC0001476 (subaward 234171B) and DE-FG02-04ER41286, which enabled the successful execution of this work.